\documentclass[acmtog]{acmart} 

\acmJournal{TOG}

\usepackage{booktabs} 
\setcitestyle{square}
\usepackage{ifthen}
\usepackage{enumitem}
\usepackage{algorithm}
\usepackage{algorithmicx}
\usepackage[noend]{algpseudocode}
\usepackage{syntax}
\usepackage{amsfonts}
\usepackage{listings}
\usepackage{fancyvrb}
\usepackage{wrapfig}
\usepackage{graphicx}
\usepackage{subcaption}
\usepackage{xspace}
\usepackage{comment}
\usepackage{hyperref}
\usepackage{multirow}

\setlist{leftmargin=*} 

\algdef{SE}[DOWHILE]{Do}{doWhile}{\algorithmicdo}[1]{\algorithmicwhile\ #1}%

\usepackage{algorithm}  
\usepackage{algpseudocode}  
\usepackage{amsmath}  

\usepackage{colortbl}
\makeatletter
\newcommand{\oset}[3][0ex]{%
  \mathrel{\mathop{#3}\limits^{
    \vbox to#1{\kern-2\ex@
    \hbox{$\scriptstyle#2$}\vss}}}}
\makeatother

\definecolor{mygreen}{rgb}{0,0.6,0}                                             
\definecolor{mygray}{rgb}{0.5,0.5,0.5}                                          
\definecolor{codebg}{rgb}{1, 1, 0.9}  
\usepackage[normalem]{ulem}

\renewcommand{\grammarlabel}[2]{#1\hfill#2}

\begin{document}
\title{Co-Optimization of Design and Fabrication Plans for Carpentry}

\author{Haisen Zhao}
\email{haisen@cs.washington.edu}
\affiliation{%
  \institution{University of Washington and Shandong University}
}

\author{Max Willsey}
\email{mwillsey@cs.washington.edu}
\affiliation{%
  \institution{University of Washington}
}

\author{Amy Zhu}
\email{amyzhu@cs.washington.edu}
\affiliation{%
  \institution{University of Washington}
}

\author{Chandrakana Nandi}
\email{cnandi@cs.washington.edu}
\affiliation{%
  \institution{University of Washington}
}

\author{Zachary Tatlock}
\email{ztatlock@cs.washington.edu}
\affiliation{%
  \institution{University of Washington}
}

\author{Justin Solomon}
\email{jsolomon@mit.edu}
\affiliation{%
  \institution{Massachusetts Institute of Technology}
}

\author{Adriana Schulz}
\email{adriana@cs.washington.edu}
\affiliation{%
  \institution{University of Washington}
}

\begin{abstract}

Past work on optimizing fabrication plans given a carpentry design
  can provide  Pareto-optimal plans trading off between
  material waste, fabrication time, precision, and other considerations.
However, when developing fabrication plans, experts rarely restrict to a \textit{single design}, instead
  considering \textit{families of design variations}, sometimes adjusting designs to simplify fabrication. Jointly exploring the design 
  and fabrication plan spaces for each design
  is intractable using current techniques.
We present a new approach to
  jointly optimize design and fabrication plans 
  for carpentered objects.
To make this bi-level optimization tractable,
  we adapt recent work from program synthesis 
  based on equality graphs (\egraphs), which encode
   sets of equivalent programs.
Our insight is that subproblems within
  our bi-level problem share significant substructures.
By representing both designs and fabrication plans
  in a new \textit{bag of parts} (BOP) \egraph,
  we amortize the cost of optimizing design components
  shared among multiple candidates.
Even using BOP \egraphs,
  the optimization space
  grows quickly in practice.
Hence,
  we also show how a feedback-guided search strategy
  dubbed \textit{Iterative Contraction and Expansion on E-graphs} (ICEE)
  can keep the size of the \egraph manageable
  and direct the search toward promising candidates.
We illustrate the advantages of our pipeline
  through examples from the carpentry domain.

\end{abstract}

%
%




\begin{teaserfigure}
\vspace{-10pt}
  {\phantomsubcaption\label{fig:teaser-a}}
  {\phantomsubcaption\label{fig:teaser-b}}
  \includegraphics[width=\textwidth]{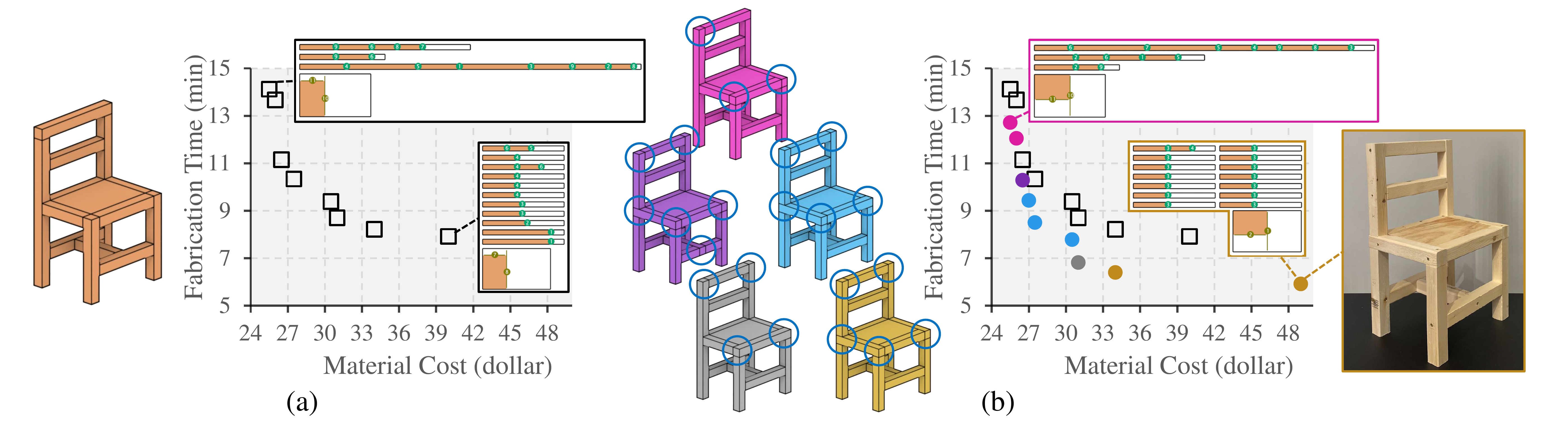}
	\vspace{-20pt}
  \caption{
  Our system jointly explores
  the space of discrete design variants and fabrication plans to generate
  a Pareto front of (design, fabrication plan) pairs that minimize
  fabrication cost.
  \change{
  In this figure,
  (a) is the input design for a chair and the Pareto front that only explores the space of
  fabrication plans for this design,
  (b) shows the Pareto front generated by joint exploration of both the
  design variants and fabrication plans for the chair, where each point is a (design, fabrication plan) pair.
  Design variations indicate different ways to compose the same 3D model from a collection of parts and are illustrated with the same color in the Pareto front.
  A physical chair is fabricated by following the result fabrication plan.
  This example shows that the fabrication cost can be significantly improved by exploring design variations.
  }
  }
  \label{fig:teaser}
\end{teaserfigure}

\newcommand{\hsyntax}[1]{\ensuremath{\mathit{#1}}}
\renewcommand{\grammarlabel}[2]{#1\hfill#2}

\newcommand{\adriana}[1]{{\bfseries \scriptsize \textcolor[rgb]{0.00,0.70,0.00}{AS: #1}}}
\newcommand{\zach}[1]{{\bfseries \scriptsize \color{red} ZT: #1}}
\newcommand{\justin}[1]{{\bfseries \scriptsize \color{blue} JS: #1}}
\newcommand{\amy}[1]{{\bfseries \scriptsize \textcolor[rgb]{0.80,0.00,0.80} {AZ: #1}}}
\newcommand{\haisen}[1]{{\bfseries \scriptsize \color{red} HZ: #1}}
\newcommand{\chandra}[1]{{\bfseries \scriptsize \color{purple} CN: #1}}
\renewcommand{\max}[1]{{\bfseries \scriptsize \color{orange} MW: #1}}

\newcommand{\change}[1]{{\color{black}#1}}

\newcommand{\Egraph}{{E-graph}\xspace}
\newcommand{\Egraphs}{{E-graphs}\xspace}
\newcommand{\egraph}{{e-graph}\xspace}
\newcommand{\egraphs}{{e-graphs}\xspace}
\newcommand{\eclass}{{e-class}\xspace}
\newcommand{\eclasses}{{e-classes}\xspace}
\newcommand{\Eclasses}{{E-classes}\xspace}
\newcommand{\enode}{{e-node}\xspace}
\newcommand{\enodes}{{e-nodes}\xspace}
\newcommand{\Enodes}{{E-nodes}\xspace}
\newcommand{\llhelm}{{LL-HELM}\xspace}
\newcommand{\hlhelm}{{HL-HELM}\xspace}
\newcommand{\fabplan}{{fabrication plan}\xspace}
\newcommand{\fabplans}{{fabrication plans}\xspace}
\newcommand{\tool}{{Carpentry Compiler 2.0}\xspace}
\newcommand{\submodule}{{e-class}\xspace}
\newcommand{\submodules}{{e-classes}\xspace}
\newcommand{\subprogram}{{sub-program}\xspace}
\newcommand{\subprograms}{{sub-programs}\xspace}
\newcommand{\afabvar}{{fabrication arrangement}\xspace}
\newcommand{\afabvars}{{fabrication arrangements}\xspace}
\newcommand{\cutEplan}{{cutting order related plan}\xspace}
\newcommand{\cutEplans}{{cutting order related plans}\xspace}
\newcommand{\term}{\ensuremath{\mathcal{T}}}
\newcommand{\bope}{{BOP E-graph}\xspace}
\newcommand{\uscore}{\ensuremath{I_{score}}\xspace}
\newcommand{\escore}{\ensuremath{E_{score}}\xspace}
\newcommand{\sols}{\ensuremath{\mathcal{S}}\xspace}
\newcommand{\nd}{\ensuremath{|\mathcal{D}|}\xspace}

\newcommand{\fabheur}{Heuristic Driven Fabrication Variations\xspace}

\maketitle
\section{Introduction}

While optimizing designs for fabrication is a long-standing and well studied engineering problem, 
 the vast majority of the work in this area assumes that there is a unique map from a design to a \fabplan.
In reality, however, many applications allow for \emph{multiple fabrication alternatives}. 
Consider, for example, the model shown in \autoref{fig:teaser-a}
 where different \fabplans trade off material cost and fabrication time.
In this context, fabrication-oriented design optimization becomes even more challenging,
 since it requires exploring the landscape of optimal \fabplans for \emph{many} design variations.
Every variation of the original design (\autoref{fig:teaser-b}) 
 determines a new landscape of \fabplans with different cost trade-offs.
Designers must therefore navigate the \emph{joint} space of design and 
 fabrication plans to find the optimal landscape of solutions.

In this work, we present a novel approach 
 that simultaneously optimizes both the design and fabrication plans for carpentry.
Prior work
 represents carpentry designs and \fabplans as programs \cite{wu2019carpentry}
 to optimize the \fabplan of a \emph{single design} at a time.
Our approach also uses a program-like representation,
 but we \emph{jointly} optimize the design and the \fabplan.

Our problem setting has two main challenges.
First, the discrete space of \fabplan alternatives 
 can vary significantly for each discrete design variation.
This setup can be understood as a \emph{bi-level} problem,
 characterized by the existence of two optimization problems 
 in which the constraint region of the upper-level problem (the joint space of designs and \fabplans)
 is implicitly determined by the lower-level optimization problem (the space of feasible \fabplans given a design).
The second challenge is that there are multiple conflicting fabrication objectives.
Plans that improve the total production time may waste more material or involve less precise cutting operations.
Our goal is therefore to find  \emph{multiple} solutions to our fabrication problem that represent optimal points in the landscape of possible trade-offs, called the \emph{Pareto front}.
Importantly, the different \fabplans on the Pareto front may come from different design variations.
The complexity of the bi-level search space combined with the need for finding a landscape of Pareto-optimal solutions makes this optimization challenging.


 


We propose a method to make this problem computationally tractable in light of the challenges above.
Our key observation is that there is redundancy on both levels of the search space that can be exploited. 
In particular, different design variations may share similar subsets of parts, which can use the same fabrication plans. 
We propose exploiting this sharing to encode a large number of design variations and their possible \fabplans compactly.
We use a data structure called an  \emph{equivalence graph (e-graph)}~\cite{egraphs} to 
 maximize sharing and thus
 amortize the cost of heavily optimizing part of a design since
 all other design variations sharing a part benefit from its optimization.

\Egraphs have been growing in popularity in the programming languages community;
 they provide a compact representation for equivalent programs
 that can be leveraged for theorem proving and code optimization. 
There are two challenges in directly applying \egraphs 
 to design optimization under fabrication variations, detailed below.


First, the different fabrication plans for a given design are all semantically equivalent programs. 
However, the fabrication plans associated with different design variations, 
 in general, are \textit{not semantically equivalent}, 
 i.e., they may produce different sets of parts.
This makes it difficult to directly apply traditional techniques which exploit sharing by searching for minimal cost, but still semantically equivalent, versions of a program. 
One of our key technical contributions is therefore a new data structure for representing the search space, which we call the \textbf{Bag-of-Parts (BOP) E-graph}. This data structure takes advantage of common substructures across both design \emph{and} \fabplans to maximize redundancy and boost the expressive power of e-graphs. 
Second, optimization techniques built around e-graphs have adopted a two stage approach: \emph{expansion} (incrementally growing the e-graph by including more equivalent programs\footnote{In the programming languages literature, this is known as \emph{equality saturation}.}) followed by \emph{extraction} (the process of searching the e-graph for an optimal program). In particular, the expansion stage has not been feedback-directed, i.e., the cost of candidate programs has only been used in extraction, but that information has not been fed back in to guide further e-graph expansion.
A key contribution of our work is 
a method for \textbf{Iterative Contraction and Expansion on E-graphs (ICEE)}.
Because ICEE is feedback-directed, it enables us to effectively explore the large combinatorial space of designs and their corresponding fabrication plans.
ICEE also uses feedback to prune the least valuable parts of the e-graph during search, keeping its size manageable.
Further, these expansion and contraction decisions are driven by a multi-objective problem that enables finding a diverse set of points on the Pareto front.

We implemented our approach and compared it against prior work
and against results generated by carpentry experts.
Our results show that
ICEE is up to $17 \times$ faster than prior approaches
while achieving similar results.
In some cases, it is the only approach that successfully
generates an optimal set of results due to its
efficiency in exploring large design spaces.
\change{
We showcase how our method can be applied to a variety of designs of different complexity and show how our method is advantageous in diverse contexts. For example we achieve 25\% reduced material in one model, 60\% reduced time in another, and 20\% saved total cost in a third when assuming a carpenter charges \$40/h, when compared to a method that does not explore design variations.
}


\section{Related Work}

\paragraph{Optimization for Design and Fabrication}


Design for fabrication is an exciting area of research that aims to automatically achieve desired properties while optimizing fabrication plans. Examples of recent work include computational design of glass fa{\c{c}}ades~\cite{gavriil2020computational}, compliant mechanical systems~\cite{tang2020harmonic}, barcode embeddings~\cite{maia2019layercode}, and \change{interlocking assemblies~\cite{wang2019design, cignoni2014field, hildebrand2013orthogonal}, among many others~\cite{bickel:surveyStylizedFab, schwartzburg2013fabrication}}.
Fabrication considerations are typically taken into account as constraints during design optimization, but these methods assume that there is an algorithm for generating \emph{one} \fabplan for a given design.
To the best of our knowledge, no prior work explores the multi-objective space of fabrication alternatives during design optimization.


There is also significant literature on \fabplan optimization for a \emph{given} design under different constraints. Recent work includes optimization of composite molds for casting~\cite{alderighi2019volume}, tool paths for 3D printing~\cite{zhao2016connected,etienne2019curvislicer}, and decomposition for CNC milling~\cite{mahdavi2020vdac,yang2020dhfslicer}. 
While some of these methods minimize the distance to a target design under fabrication constraints~\cite{zhang2019computational,duenser2020robocut}, none of them explores a space of design modification to minimize fabrication cost. 

%
%
In contrast, our work \emph{jointly} explores the design and fabrication space in the carpentry domain, searching for the Pareto-optimal design variations that minimize multiple fabrication costs.

\paragraph{Design and Fabrication for Carpentry}

Carpentry is a well-studied domain in design and fabrication due to its wide application scope. Prior work has investigated interactive and optimization methods for carpentry design~\cite{Umetani:GuidedExploration,koo2014creating,Song2017reconfigurableFurniture, garg2016computational,Fu:InterlockingFurniture}. There is also a body of work on \fabplan  optimization~\cite{yang2015reforming,Koo:ZeroWasteFurniture,jigfab,Lau2011FabPartsConnectors}. Closest to our work is the system of \citet{wu2019carpentry}, which represents both carpentry designs and \fabplans as programs and introduces a compiler that optimizes \emph{fabrication} plans for a \emph{single} design. While our work builds on the domain specific languages (DSLs) proposed in that prior work, ours is centered on the fundamental problem of design optimization under fabrication alternatives, which has not been previously addressed.

\paragraph{Bi-Level Multi-Objective Optimization}

Our problem and others like it are \emph{bi-level}, with a nested structure in which each design determines a different space of feasible \fabplans. 
The greatest challenge in handling bi-level problems lies in the fact that the lower level problem determines the feasible space of the upper level optimization problem. More background on bi-level optimization can be found in the book by~\citet{dempe2018bilevel}, as well as review papers by~\citet{lu2016multilevel} and~\citet{sinha2017review}.

Bi-level problems with multiple objectives can be even more challenging to solve~\cite{dempe2018bilevel}. Some specific cases are solved with classical approaches, such as numerical optimization~\cite{eichfelder2010multiobjective} and the $\epsilon$-constraint method~\cite{shi2001model}. 
Heuristic-driven search techniques have been used to address bi-level multi-objective problems, such as genetic algorithms~\cite{yin2000genetic} and particle swarm optimization~\cite{halter2006bilevel}. These methods apply a heuristic search to both levels in a nested manner, searching over the upper level with NSGA-II operations, while the evaluating each individual call in a low-level NSGA-II process~\cite{deb2009solving}. Our ICEE framework also applies a genetic algorithm during search. Different from past techniques, ICEE does not nest the two-level search but rather reuses structure between different upper-level feasible points. ICEE jointly explores both the design and fabrication spaces using the \bope representation.

\paragraph{E-graphs}

An \egraph is an efficient data structure
for compactly representing large sets of equivalent programs.
\Egraphs were originally developed for
automated theorem proving~\citep{egraphs}, and
were first adapted for program optimization by~\citet{denali}.
These ideas were further expanded to handle programs with loops and
conditionals~\cite{eqsat} and applied to a variety of domains for program
optimization, synthesis, and equivalence checking~\citep{
    eqsat-llvm, egg, szalinski, herbie, wu2019carpentry, spores, yogo}.

Recently, \egraphs have been used for optimizing designs~\cite{szalinski}, and
also for optimizing fabrication plans~\cite{wu2019carpentry}, but
they have not been used to simultaneously optimize \textit{both} designs
and fabrication plans.
Prior work also does not explore
feedback-driven \egraph expansion and contraction
for managing large optimization search spaces.

\section{Background}
In this section, we introduce some mathematical preliminaries used in the rest of the paper.

\subsection{Multi-Objective Optimization}
\label{mathprelim:moo}

A multi-objective optimization problem is defined by set of objectives $f_i:\mathbf{x} \mapsto \mathbb{R}$ that assign a real value to each point $\mathbf{x} \in \mathcal{X}$ in the feasible search space $\mathcal X$.  We choose the convention that \emph{small} values of $f_i(\mathbf{x})$ are desirable for objective $f_i$. 

As these objectives as typically \emph{conflicting}, our algorithm searches for a diverse set of points that represent optimal trade-offs, called \change{ \emph{Pareto optimal}~\citep{deb2014multi}}:

\begin{definition}[Pareto optimality]
\label{def:paretoOpt}
A point $\mathbf{x}\in\mathcal X$ is \emph{Pareto optimal} if there does not exist any $\mathbf{x}'\in\mathcal X$ so that $f_i(\mathbf{x})\geq f_i(\mathbf{x}')$ for all $i$ and $f_i(\mathbf{x})> f_i(\mathbf{x}')$ for at least one $i$.
\end{definition}

We use $F:\mathbf{x} \mapsto\mathbb{R}^N$ to denote the concatenation $(f_1(\mathbf x),\ldots,f_N(\mathbf x))$. Pareto optimal points are the solution to the multi-objective optimization:
\begin{equation}\label{eq:primal}
\min_{\mathbf x}  F(\mathbf x) \ \  \mathrm{s.t.} \  \mathbf x \in\mathcal X.
\end{equation}
The image of all Pareto-optimal points is called the \emph{Pareto front}.

\paragraph{Non-Dominated Sorting}
Genetic algorithms based on non-dominated sorting are a classic approach to multi-objective optimization~\cite{Deb:2002:FEM:2221359.2221582,deb2013evolutionary}. The key idea is that sorting should be done based on proximity to the Pareto front. These papers define the concept of Pareto layers, where layer $0$ is the Pareto front, and layer $l$ is the Pareto front that would result if all solutions from layers $0$ to $l-1$ are removed. When selecting parent populations or when pruning children populations, solutions in lower layers are added first, and when a layer can only be added partially, elements of this layer are chosen to increase diversity. Different variations of this method use different strategies for diversity; we use NSGA-III~\cite{deb2013evolutionary} in our work.  

\paragraph{Hypervolume}
\textit{Hypervolume} ~\cite{auger2009theory} is a metric commonly used to compare two sets of image points during Pareto front discovery. 
To calculate the hypervolume, we draw the smallest rectangular prism (axis-aligned, as per the $L^1$ norm) 
between some reference point and each point on the pareto front.
We then union the volume of each shape to calculate the hypervolume.
Thus, a larger hypervolume implies a better approximation of the Pareto front.

\subsection{Bi-level Multi-Objective Optimization}
\label{mathprelim:bilevel-moo}
 Given a design space $\mathcal{D}$ that defines possible variations of a carpentry model, our goal is to find a design $d \in \mathcal{D}$ and a corresponding \fabplan $p\in \mathcal P^d$ that minimizes a vector of conflicting objectives, where $\mathcal P^d$ is the space of \fabplans corresponding to design $d$. This setup yields the following multi-objective optimization problem: 

\begin{equation*}
  \min_{p,d} F(d, p) \ \   \text{s.t.} \ \ d \in \mathcal{D}, \ \ p \in \mathcal{P}^d
\end{equation*}
where $\mathcal{P}^d$ defines the space of all possible plans for fabrication the design $d$. 
Generally, our problem can be expressed as a bi-level multi-objective optimization \change{that searches across designs to find those with the best fabrication costs, and requires optimizing the fabrication for each design during this exploration}~\cite{lu2016multilevel}:
\begin{equation*}
   \min_{d} F(d, p)  \ \  \text{s.t.} \ \ \ \ d \in \mathcal{D},  \ \ \
 p = \arg \min_{p} F(d, p)     
\end{equation*}
where $\arg\min$ refers to Pareto-optimal        solutions to the multi-objective optimization problem. 

A na\"ive solution to this bi-level problem would be to search over the design space $\mathcal{D}$ using a standard multi-objective optimization method, while solving the nested optimization problem to find the \fabplans given a design at each iteration. Given the combinatorial nature of our domain, this would be prohibitively slow, which motivates our proposed solution.



\subsection{Equivalence Graphs (E-graphs)}
\label{mathprelim:egraphs}
\label{subsec:egraphs}

Typically, programs (often referred to as \textit{terms}) 
 are viewed as tree-like structures containing smaller sub-terms.
For example, 
 the term $3 \times 2$ has
 the operator $\times$ at its ``root'' and
 two sub-terms, $3$ and $2$, 
 each of which has no sub-terms.
Terms
 can be expressed in multiple syntactically different ways.
For example, 
 in the language of arithmetic,
 the term $3 \times 2$ is semantically equivalent to $3+3$, 
 but they are syntactically different.
Na\"ively computing and storing all semantically equivalent
 but syntactically different variants of the
 a term requires exponential time and memory.
For a large program, 
 this makes searching the space of equivalent terms intractable.

\textit{\Egraphs}~\cite{egraphs} are designed to address this challenge---an 
 \egraph is a data structure
 that represents many equivalent terms efficiently
 by sharing sub-terms whenever possible.
An \egraph not only stores a large set of terms,
 but it represents an \textit{equivalence relation}
 over those terms,
 i.e.,
 it partitions the set of terms into 
 \textit{equivalence classes}, or \textit{\eclasses},
 each of which contains semantically equivalent 
 but syntactically distinct terms.
\change{
In \autoref{subsec:bop-egraph},
 we show how to express carpentry designs
 in a way that captures the benefits of the \egraph.
}

\begin{figure}[t]
  \includegraphics[width=40mm]{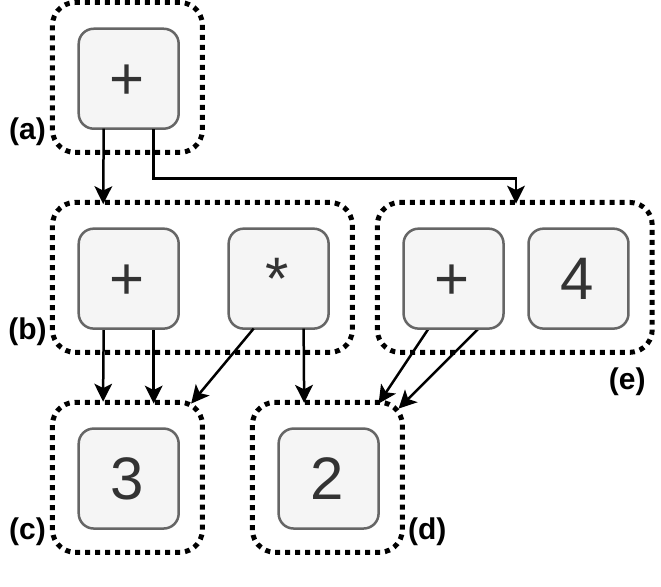}
  \newcommand{\ecref}[1]{\textsf{\textbf{(#1)}}}
  \caption{
    An example \egraph.
    \Eclasses (dotted boxes labelled by letters) contain equivalent \enodes (solid boxed)
     which refer to children \eclasses (arrows).
    The \eclass \ecref{c} contains one leaf \enode, $3$, 
     and it represents one term, $3$.
    The \eclass \ecref{b} contains two \enodes, $\ecref{c} + \ecref{c}$ and $\ecref{c} * \ecref{d}$,
     and it represents two terms: $3 + 3$ and $3 * 2$.
    \change{
    Although the \eclass \ecref{a} only contains one \enode,
     it represents 4 terms:
     $(3 + 3) + (2 + 2)$,
     $(3 * 2) + (2 + 2)$,
     $(3 + 3) + 4$,
     and
     $(3 * 2) + 4$.
    If $+$ is cheaper than $*$,
     then 
     $(3 + 3) + 4$ is the cheapest term represented by \eclass \ecref{a}.
    }
  }
  \label{fig:eg-overview}
\end{figure}

\begin{definition}[\Egraph]
\label{def:egraph}
An \textit{\egraph} is a set of equivalence classes or
\textit{\eclasses}.
An \textit{\eclass} is a set of equivalent \enodes.
An \textit{\enode} is
 an operator from the given language paired with some \eclass children,
 i.e.,
 $f(c_1,\ldots,c_n)$ is an \enode where $f$ is an operator and each $c_i$ is an \eclass that is a child of this \enode.
An \enode may have no children, in which case we call it a \textit{leaf}.
An \egraph \textit{represents} an equivalence relation over terms.
Representation is defined recursively:
\begin{itemize}
    \item An \egraph represents a term if any of its \eclasses do.
    \item An \eclass represents a term if any of its \enodes do.
          All terms represented by \enodes in the same \eclass are equivalent.
    \item An \enode $f(c_1,\ldots,c_n)$ represents a term $f(t_1,\ldots,t_n)$ 
          if each \eclass $c_i$ represents term $t_i$.  
          A leaf \enode $g$ represents just that term $g$.
\end{itemize}

\autoref{fig:eg-overview} shows an example of an \egraph and representation.
Note how the \egraph maximizes sharing even across syntactically distinct, 
 semantically equivalent terms.
When adding \enodes or combining \eclasses, 
 the \egraph automatically maintains this maximal sharing property,
 using existing \enodes whenever possible.

\end{definition}

\section{Optimization Algorithm}

\begin{figure}[h!]
    \centering
    \includegraphics[width=\linewidth]{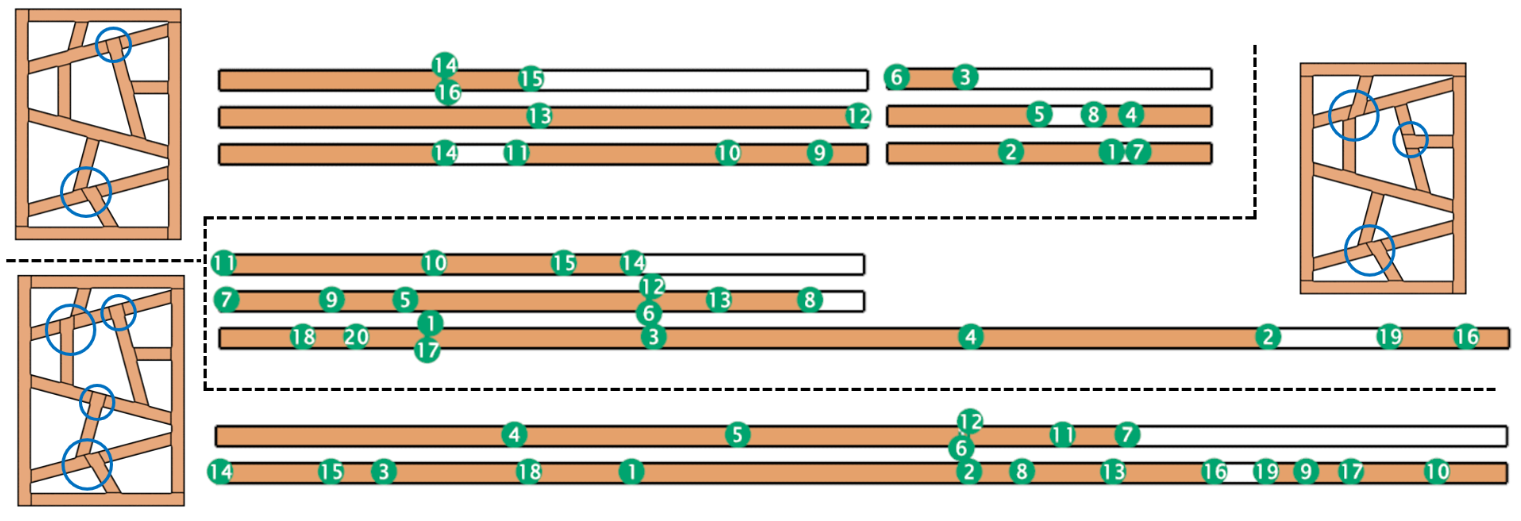}
    \caption{\change{Example of three different design variations of a model and corresponding fabrication plans. Design variations determine different ways to decompose a 3D model into a set of parts. Fabrication plans define how these parts are \emph{arranged} in pieces of stock material and the cut order (illustrated by the numbers along each cut). }}
    \label{fig:exampledefinitions}
\end{figure}

Our algorithm takes as input a carpentry design with a discrete set $\mathcal{D}$
 of possible design variations. \change{Design variations determine different ways to decompose a 3D model into a set of fabricable parts, as shown in Figures~\ref{fig:exampledefinitions} and \ref{fig:joints}.}
These can be manually or automatically generated (\change{see Section 1.1 of the supplemental material}).

\change{
Our goal is to find Pareto-optimal solutions that minimize fabrication cost,
 where each solution is a pair of design variation and fabrication plan. 
}
\change{Similar to prior work~\citep{wu2019carpentry},
 we measure cost in terms of
 material usage ($f_c$),
 cutting precision ($f_p$), and
 fabrication time ($f_t$).
 \change{Section 1.3 of the supplemental material} describes how these metrics
 are computed for this work.}


\begin{figure}[ht]
    \centering
    \includegraphics[width =0.85\linewidth]{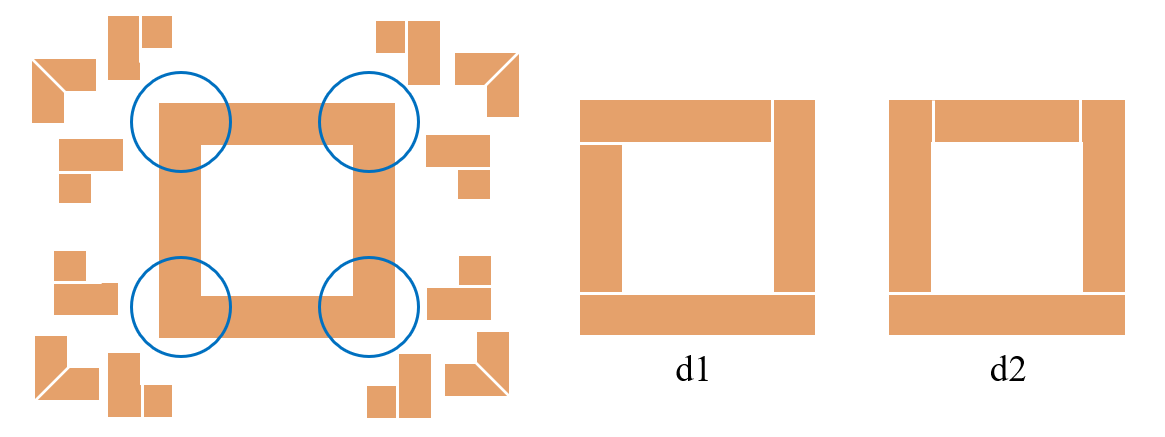}
    \caption{Example of a space of design variations, $\mathcal{D}$. Each of the four connectors can have three different connecting variations, resulting in 81 design variations. Note that some of the different design variations may use the same parts (as d1, d2), and will be treated as redundant during our optimization. This model produces 13 unique bags of parts.} 
    \label{fig:joints}
\end{figure}


\subsection{Motivation and Insights}

Given an algorithm for finding the Pareto-optimal fabrication plans for a given design
 (e.g., the one proposed by Wu et al.~\shortcite{wu2019carpentry}), 
 a brute force method would simply find the Pareto-optimal solutions for each 
 of the possible design variations $d \in \mathcal{D}$ 
 and take the dominant ones to form the Pareto front of the combined design/fabrication space.
Since design variations can produce an exponentially large space of designs $\mathcal{D}$, this approach would be intractable for complex models. 
An alternative approach could use a discrete optimization algorithm to explore the design space (e.g. hill climbing). 
This approach would still need to compute the Pareto-optimal fabrication plans for each design explored in every iteration, which can expensive for complex design variants (e.g., it takes 8-10 minutes to compute Pareto-optimal fabrication plans for a single design variation of the chair model in \autoref{fig:teaser} using the approach of Wu et al.~\shortcite{wu2019carpentry}). 

We address these challenges with two key insights: 
\begin{enumerate}
    \item 
    Design variants will share common sub-parts
     \change{(both within a single variant and across different variants). As shown in Figure~\ref{fig:exampledefinitions}, even in a design where no two parts are the same, there is significant overlap \emph{across} design variations.}
     Exploiting this sharing
     can prevent recomputing the fabrication cost from scratch for every design variation.
    We propose using a \egraph to capture 
     this sharing when (sub-)designs have the same bag of parts;
     we call this \egraph the \textit{\bope}.
    \item
    The space of design variants is too large to exhaustively explore,
     and even a single variant may have many Pareto-optimal fabrication plans.
    We propose using the \bope to guide the exploration in an incremental manner,
     with a new technique called \textit{ICEE} (Iterative Contraction and Expansion of the \Egraph)
     that jointly explores the design and fabrication plan spaces.
\end{enumerate}

\subsection{Bag of Parts (BOP) \Egraph}
\label{subsec:bop-egraph}

\change{
Our algorithm
 selects a Pareto-optimal set of fabrication plans,
 each of which will produce a design variation of the given model.
A \textit{fabrication plan} consists of four increasingly detailed things:
\begin{enumerate}
    \item A \textbf{bag of parts},  
          a bag\footnotemark{} (a.k.a. multiset) of atomic parts that compose the model.
    \item An \textbf{assignment} that maps those parts to individual pieces of stock material.
    \item A \textbf{packing} for each piece of stock 
          in the assignment that dictates \emph{how} those parts are arranged in that stock.
    \item A \textbf{cutting order} for each packing that specifies the order and the tool (chopsaw, tracksaw, etc.) used to cut the stock into the parts.
\end{enumerate}
We say that an \textbf{arrangement} is items 1-3: 
 a bag of parts assigned to and packed within pieces of stock material,
 but \emph{without cutting order decided}.
}
\footnotetext{
   A \textit{bag} or \textit{multiset} is an unordered set with multiplicity, 
   i.e. it may contain the same item multiple times. We will use the terms interchangeably.
}
We can create a language to describe arrangements;
a term in the arrangement language is one of the following:
\begin{itemize}
    \item An \textbf{atomic node} is a childless operator 
          that represents a bag
          of parts packed into a single piece of stock.
          For example, $\{\square, \square, \triangle\}_{p,b}$ maps 
          two squares and one triangle all to the same piece of stock of type $b$ using a packing $p$.
    \item A \textbf{union node} takes two child arrangements and composes them into a single arrangement.
          The following arrangement is a union node of two atomic nodes:
          ${\{\square, \square\}_{p_1,b} \cup \{\triangle\}_{p_2,b}}$.
          It packs two squares into stock of type $b$ using packing $p_1$,
          and it packs a triangle into a \textit{different piece} of stock of the same type $b$ using packing $p_2$.
\end{itemize}


To put arrangements into an \egraph, 
 we must define the notion of equivalence that the \egraph uses to
 determine which \enodes go in the same \eclass.
The more flexible this notion is 
 (i.e., the larger the equivalence relation),
 the more sharing the \egraph can capture.
 
To maximize sharing,
 we say two arrangements are equivalent if they use the 
 same \textit{bag of parts} (BOP),
 even if those parts are assigned to different stock or packed differently.
For example,  
 $\{\square, \square\}_{p_1,b}$
 is equivalent to
 $\{\square, \square\}_{p_2, c}$
 even though they use different kinds of stock,
 and
 $\{\square, \square, \triangle\}_{p_3,b}$
 is equivalent to
 $\{\square, \triangle\}_{p_4,b} \cup \{\square\}_{p_5,b}$
 even though the former uses one piece of $b$ stock and the latter uses two.
 
Given our arrangement language and the BOP notion of equivalence, 
 we can now describe the central data structure of our algorithm,
 the \textit{\bope}.
Recall from \autoref{subsec:egraphs} that \enodes within an \egraph
 have \eclass children rather than \enode children.
So, viewing our arrangement language at the \egraph level,
 union \enodes take two \eclasses as children.
All \enodes in the same \eclass are equivalent,
 i.e., they represent terms that use the same bag of parts but that 
 arrange those parts differently into stock.
 
\begin{figure}
    \centering
    {\phantomsubcaption\label{fig:bope1}}
    {\phantomsubcaption\label{fig:bope2}}
    {\phantomsubcaption\label{fig:bope3}}
    \includegraphics[width = \linewidth]{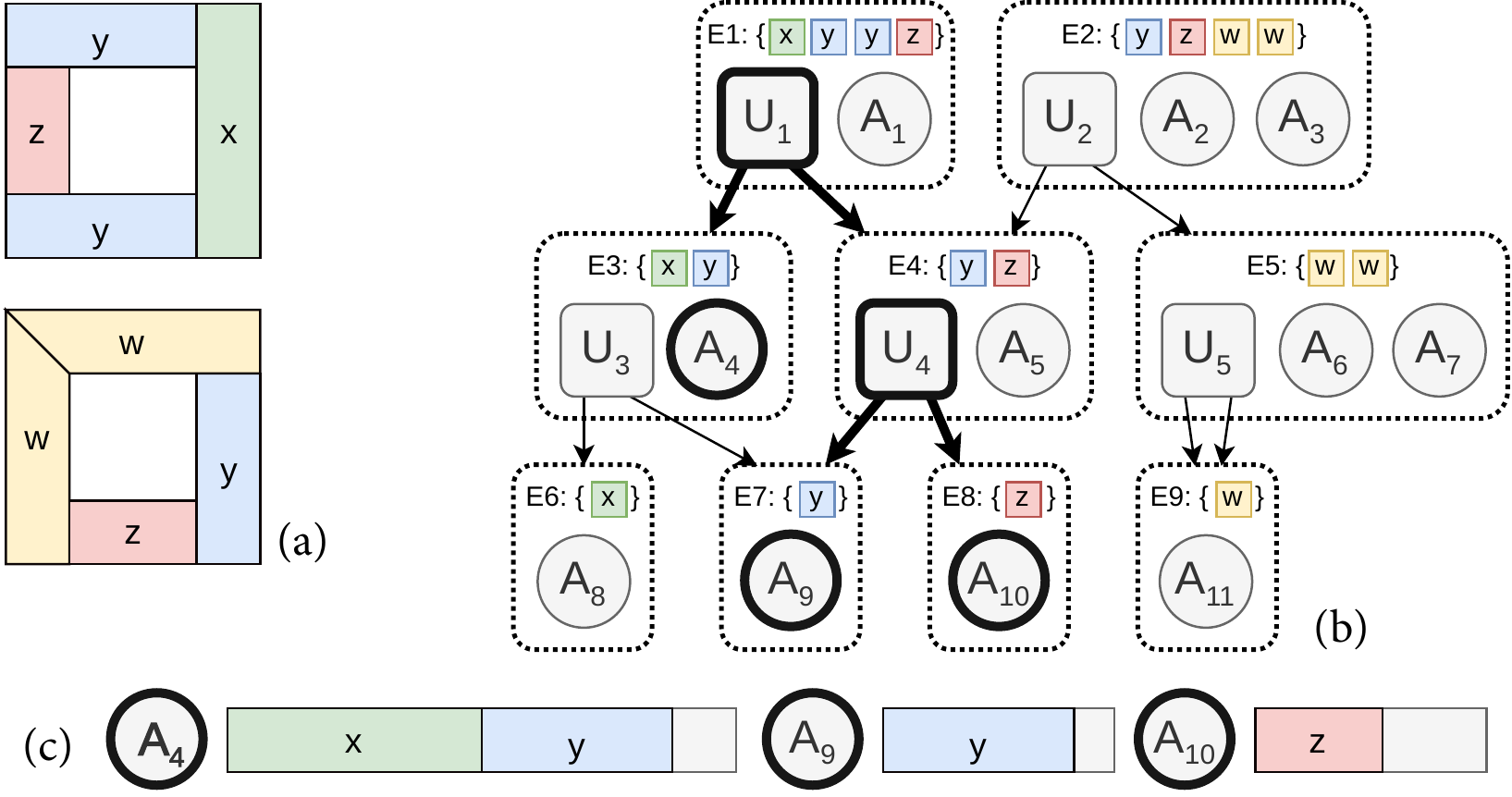}
    \caption{
    \change{
      Two variants \subref{fig:bope1} of a box design 
      encoded in one \bope \subref{fig:bope2}.
      The bold edges show a     root term
       that requires 3 atomic packings \subref{fig:bope3}.
    }
      The \bope encodes multiple arrangements for both design variants.
      \Eclasses are drawn as dotted boxes and annotated with 
       the bag of parts represented by that \eclass.
      (Only the \enodes are semantically \textit{in} the \egraph; 
       the name and bag of parts are just visual aides.)
      \Eclasses E1 and E2 are root \eclasses since they represent the bags of parts required by the design variants. 
      Union and atomic \enodes are shown as squares with ``U''s or circles with ``A''s, respectively.
      Atomic \enodes correspond to packings of parts within a piece of stock \subref{fig:bope3}.
      An example root term in the \bope is bolded;
       using the syntax from \autoref{subsec:bop-egraph}, this is the term
      $\{x, y\}_{A_4,\textsf{long}} \cup 
       \{y\}_{A_9,\textsf{short}} \cup \{z\}_{A_{10},\textsf{short}}$.
        }
    \label{fig:bope}
\end{figure}

\change{
\autoref{fig:bope} gives two example design variations and a \bope 
 that captures a common sub-arrangement between the two.
The \eclasses E1 and E2 represent terms 
 that correspond to the two box designs,
 and E4 captures ways to arrange the $y$ and $z$ parts which
 the variants share.
The design variant including part $w$ also captures sharing with itself:
 \eclass E5 reuses the arrangement in \eclass $E9$.
}

Note that arrangements and the \bope do not mention designs.
We do not ``store'' designs in the \egraph,
 we just need to remember which \eclasses represent 
 bags of parts that correspond to designs 
 that we are interested in.
This can be done outside the \egraph with a 
 mapping from designs to \eclasses.
Many designs (especially symmetric ones) 
 may have the same bag of parts.
We call an \eclass that is associated with a design 
 a \textit{root \eclass}, 
 and we call a term represented by a root \eclass a \textit{root term}.
The \bope itself does not handle root vs.\ non-root \eclasses or terms differently,
 these are only used by the remainder of the algorithm 
 to remember which arrangements correspond to design variants.
The \bope will maximize sharing across design variations \textit{and} arrangements
 since it makes no distinction between the two.

\begin{figure*}
\centering
\includegraphics[width=\linewidth]{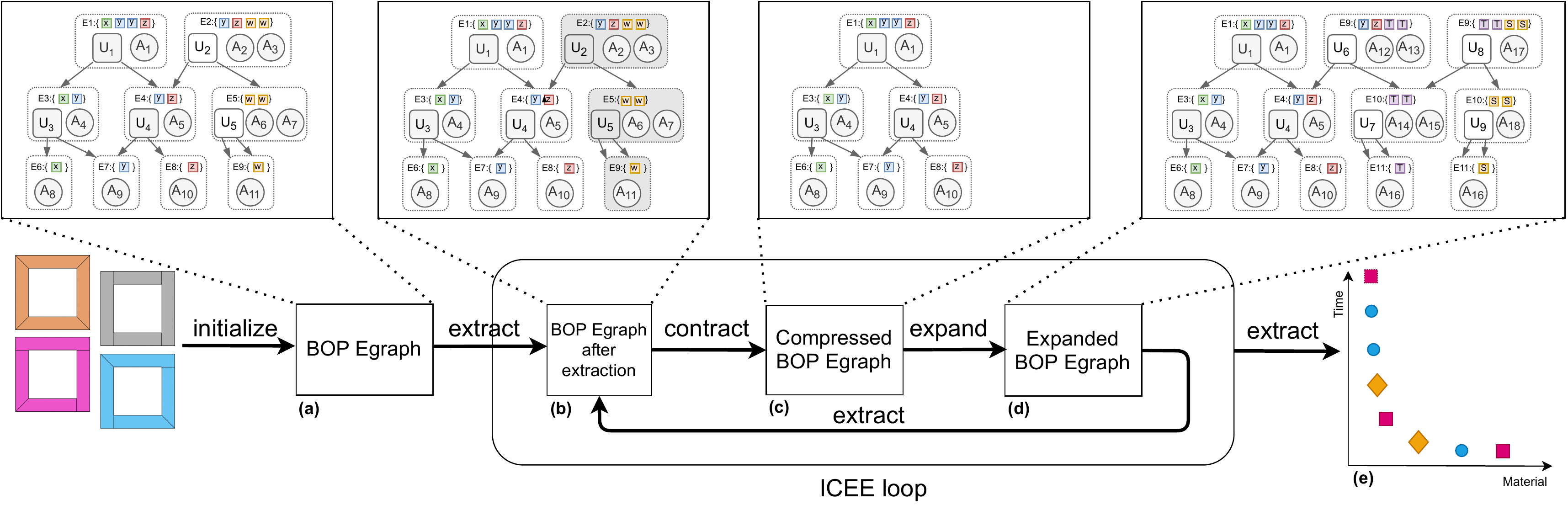}
\caption{Algorithm overview \change{used the example in~\autoref{fig:bope}}.
The first step initializes a
  \bope (\autoref{subsec:des-var}, \autoref{subsec:fabheur})
  with several design variants and
  a small number of fabrication arrangements (a).
  \texttt{U} and \texttt{A} represent union and atomic \enodes respectively.
As part of the ICEE loop,
  the algorithm extracts a Pareto Front (\autoref{subsec:extract}) which
  is used to score the \eclasses in the \bope (b).
For example, the gray \eclass containing a ``\texttt{U}'' and
an ``\texttt{A}'' \enode indicates a low score,
i.e., the \eclass did not contribute to Pareto-optimal solutions.
The \bope is then contracted (\autoref{subsec:constraction})
 by removing the low-scored \eclasses (and their parent \enodes) 
 to get a compressed \bope (c).
 As described in \autoref{subsec:extension},
 this contracted \bope is
 then further expanded (d) by exploring
  more design variants and fabrication arrangements.
The algorithm exits the loop when the termination
  conditions are reached,
  returning the final Pareto Front (e).
}
\label{figPipeline}
\label{fig:overview}
\end{figure*}

\subsection{Iterative Contraction and Extension on E-graphs (ICEE)}
\label{subsec:icee}




\subsubsection{Overview}

ICEE takes a feasible design space $\mathcal{D}$ as input, and outputs a Pareto front where each solution $s$ represents a (design, fabrication) pair. An overview of this algorithm is shown in Figure~\ref{fig:overview}.

The initialization step selects a small subset  of design variants from $\mathcal{D}$ (\autoref{subsec:des-var}) and then generates a small number of \afabvars for each one (\autoref{subsec:fabheur}). All of these are added to the \bope, maintaining the property of maximal sharing, as described above. ICEE then applies the extraction algorithm (\autoref{subsec:extract}) to generate a Pareto front from the current \bope. This process will compute many different solutions $s$ and their fabrication costs $F(s) = (f_m(s),f_p(s),f_t(s) )$, all of which are stored in the solution set $\sols$. 

The resulting Pareto front is used to compute ranking scores for each \eclass in the \bope; the ranking score measures how often this bag of parts is used in Pareto-optimal solutions and how many fabrication variations have been explored for this bag of parts. Using these scores, ICEE contracts the \bope by pruning \eclasses that have been sufficiently explored but still do not contribute to Pareto-optimal solutions (\autoref{subsec:constraction}). 

Having pruned the \egraph of the less relevant \eclasses, ICEE then expands the \bope in two ways (\autoref{subsec:extension}). First, it suggests more design variations based on the extracted Pareto-Optimal designs. Second, it  generates more \afabvars
for both the new generated design variations and some of the previously existing \eclasses. The ranking scores are used to select \eclasses for expansion.

ICEE then extracts the new Pareto front from
 the updated \bope and repeats the
 contraction and expansion steps
 until the following termination criteria are met:
 1) there is no hypervolume improvement within $t_d$ iterations, or
 2) we exceed $mt_d$ iterations.
Additionally, we set a timeout $T$ beyond which we no longer
 change the \bope, but continue to extract based on
 crossover and mutation 
 until one of the termination criteria is met.
 In our experiments, we set \change{$t_d = 10 $, $mt_d=200$}, and $T=4$  hours.

\subsubsection{Initial Generation of Design Variants}
\label{subsec:des-var}
 We bootstrap our search with the observation that design variations with more identical parts tend to be cheaper to fabricate because less time is spent setting up fabrication processes. Therefore, instead of initializing the \bope with $K_d$ designs randomly selected from $\mathcal{D}$, we randomly select \change{up to $10^5$} designs and select the top $K_d$ designs from this set that have a maximal number of identical parts.

\subsubsection{Fabrication Arrangements Generation}
\label{subsec:fabheur}
Again, instead of randomly generating $K_f$ arrangement variations for a given design, we use heuristics; namely, that (1) we can minimize the number of cuts by stacking and aligning material to cut multiple parts with a single cut, and (2) we can minimize the material cost by packing as many parts as possible to a single stock. Since a similar method for generating arrangement variations has been previously proposed by \citet{wu2019carpentry}, we leave a detailed discussion of the algorithm for supplemental material (\change{Section 1.2}). We note that the key difference between our method and the prior heuristic-driven algorithm is that we incorporate storage and direct control schemes that enable the method to output $K_f$ variations that are \emph{different} from the ones generated during previous iterations of ICEE. This is essential to enable incremental expansion of the \bope without restoring variations that have already been pruned in previous contraction steps.

\subsubsection{Pareto Front Extraction}
\label{subsec:extract}

In \egraph parlance, \textit{extraction} is the process of selecting the ``best''
 represented term from an \egraph according to some (typically single-objective) cost function.
One way to view extraction is that it simply chooses 
 which \enode should be the canonical representative of each \eclass;
 once that is done, each \eclass represents a single term.
Since our cost function is multi-objective, 
 we must instead extract a set of terms (arrangements) 
 from the \bope that forms a Pareto front.
 
We use a genetic algorithm~\cite{deb2013evolutionary} to 
 extract terms from the \bope. \change{The population size is set to $N_{pop}$.}
The genome is essentially a list of integers, 
 one per \eclass,
 that specifies which \enode is the representative.
Since the \bope may have multiple root \eclasses 
 (corresponding to multiple design variations),
 we combine the genes for all the root \eclasses, 
 only picking a single \enode among all of them.
In effect, this means the genome defines both a design variation and
 the arrangement for that design.
 
For example, consider the bold term within the \bope in \autoref{fig:bope}.
The genome for that term is as follows, where $*$ could be any integer
 since that \eclass is not used by the term:
\vspace{-0mm}$$\begin{array}{cccccccc}
     E_1, E_2 & E_3 & E_4 & E_5 & E_6 & E_7 & E_8 & E_9  \\
     0        & 1   & 0   & *   & *   & 0   & 0   & *
\end{array}$$
The root \eclasses $E_1$ and $E_2$ share a single integer $0$, 
 meaning that the genome chooses the $0$th \enode \emph{across both} \eclasses,
 and that it uses the first of the two design variants.
Since this encoding boils down to a list of integers,
 which is valid as long as each integer corresponds to an \enode in that \eclass,
 we can use simple mutation and single-point crossover operations.

A term does not completely determine a fabrication plan; it only specifies the arrangement. 
We need to additionally compute the cutting order for a given term to define a solution $s$ and then evaluate the fabrication costs.
We observe that the material cost does not depend on the cutting order and that precision and fabrication costs strongly correlate once the arrangement is fixed. 
This is not surprising since cutting orders that minimize set-ups will jointly reduce time and precision error. 
Given this observation, we can compute two solutions for each term, using two single-objective optimizations for computing cutting order: one that minimizes precision, and the other fabrication time.  

We use two strategies to speed up these optimizations: (1) storing computed cutting orders in atomic \enodes that will be shared across many terms and (2) a branch and bound technique. The optimization works as follows. Given a term, we first compute the optimal plans for the atomic \enodes that have not been previously optimized. \change{For each such \enode, we try to generate maximal $P$ different orders of cuts, then extract the optimal plans with~\cite{wu2019carpentry} method.}
We use this result to compute an upper and a lower bound for the term. If the lower bound is not dominated by the Pareto front of all computed solutions $\sols$, we run an optimization that uses the upper bound as a starting point (see Section 1.4 of the supplemental material for details).

 We again terminate the algorithm if there is no hypervolume improvement within $t_p$ iterations,
  or if we exceed $mt_p$ iterations. 
 In our experiments, we set \change{$t_p = 20 $ and $mt_p=200$} and set the probability of crossover \change{($mc_p$)} and mutation \change{($mm_p$)} are set to be $0.95$, $0.8$ respectively.

\subsubsection{BOP \Egraph Contraction}
\label{subsec:constraction}
As the algorithm proceeds, 
 \bope contraction keeps the data structure from growing too large.
To contract the \bope, we search for \eclasses that represent bags of parts that have been sufficiently explored by the algorithm but are not present in Pareto-optimal designs.
This indicates that we have already discovered the best way to fabricate these bags of parts
 but they still do not contribute to Pareto optimal solutions;
 these \eclasses are then deleted.

To measure how much an \eclass has been explored, we first compute how many variations of \afabvars have been encoded in the \bope. This number is stored over the \egraph and updated after each expansion step to ensure consistency following contraction steps. The exploration score, $\escore$, is then defined as this value divided by the number of possible \afabvars for an \eclass, which we approximate by the number of parts in the \eclass multiplied by the number of orientations of each part that can be assigned to the stock lumber. 

The impact of an \eclass, $\uscore$, is measured based on how often it is used in the set of solutions in the current Pareto front. 
We use the assignment of solutions $s$ to layers determined by the non-dominated sorting (\ref{mathprelim:moo}) to compute $\uscore$ for a given \eclass. We initialize a $\uscore$ with value $0$ and increment it by $10^{M-l}$ every time this \eclass is used in a solution from layer $l$, where $M$ is the total number of valid layers.

We normalize all computed exploration and impact scores to be between zero and one and then assign the following pruning score to each e-class:
\begin{equation*}
  P_{score} = w \cdot \uscore + (1-w) \cdot  (1-\escore), w \in [0.0,1.0]
\end{equation*}
where the weight $w$ is chosen to trade-off between exploration and impact. 
If the $P_{score}$ is smaller that the pruning rate, $P_{rate}$, the \eclass is removed along with any \enodes pointing to this \eclass (i.e. parent \enodes). We set $w$ and $P_{rate}$ to $0.7$ and $0.3$ in our implementation. 

\subsubsection{BOP \Egraph Expansion}
\label{subsec:extension}
We expand the \bope by first generating new design variations and then by generating \afabvars for both the existing and newly generated design.

We generate new design variations using a single step of a genetic algorithm that operates over the design space.
The probability of crossover \change{($mc_d$)} and mutation \change{($mm_d$)} are set to be $0.95$, $0.8$ respectively.
We select the parent design variations from $\sols$ based on the non-dominated sorting technique (\autoref{mathprelim:moo}).
Since many solutions in $\sols$ can correspond to the same design, we assign designs to the lowest layer that includes that design.
We then generate new design variations with crossover and mutation operations.
We use an integer vector encoding for each design.
This time, the vector indexes the joint variations, e.g., for the designs shown in \autoref{fig:joints}, $d_1 =[0,2,1,0], d_2 =[1,0,0,2]$.
\change{
We get $K_m \cdot K_d$ design variations by applying $K_m$ times of the single step genetic algorithm. Then we apply the same heuristic done during initialization (\autoref{subsec:des-var}), selecting the top $K_{nd}, K_{nd} \in [0, k_d]$. Finally, the resulting $K_{nd}$ designs are included to the \bope. We set $K_m =10$ in our implementation. 
}


We generate \afabvars for each of the new design variations using the algorithm described in \autoref{subsec:fabheur},
 and they are added to the \bope  maintaining the maximal sharing property.
We further generate \afabvars for existing design variations, using a similar scoring function used during contraction. This is done in two steps. First we select root \eclasses to expand based only on their impact score; namely, we take the top $K_d$ root \eclasses using non-dominated sorting. We then proceed to generate $K_f \times K_d$ \afabvars using the algorithm described in \autoref{subsec:fabheur}). However, instead generating the same number of \afabvars variations for every selected root \eclass, the number is adaptive to their pruning scores $P_{score}$ (as defined in \autoref{subsec:constraction}).

\begin{table}[]
\scriptsize
    \centering
    \begin{tabular}{|c|c|c|c|c||c|c|c|c|c|}
    \hline
     \textbf{Model} &  \textbf{$n_p$} &  \textbf{\#C} & \textbf{\#CV} &  \textbf{$\nd$} &
     \textbf{Model} &  \textbf{$n_p$} &  \textbf{\#C} &  \textbf{\#CV} &  \textbf{$\nd$}  \\ \hline
Frame      & 4  & 4  & 22 & 13    & A-Chair  & 18 & 3  & 6  & 4     \\ \hline
L-Frame    & 6  & 8  & 16 & 65    & F-Pot  & 8  & 1  & 4  & 4     \\ \hline
A-Bookcase & 12 & 6  & 16 & 192   & Z-Table  & 15 & 6  & 16 & 63    \\ \hline
S-Chair      & 14 & 14 & 32 & 66438 & Loom     & 18 & 4  & 10 & 36    \\ \hline
Table      & 12 & 10 & 24 & 1140  & J-Gym    & 23 & 8  & 16 & 54    \\ \hline
F-Cube     & 12 & 8  & 23 & 5     & D-Chair  & 17 & 10 & 22 & 2280  \\ \hline
Window     & 12 & 16 & 32 & 10463 & Bookcase & 15 & 22 & 44 & 65536 \\ \hline
Bench      & 29 & 6  & 14 & 57    & Dresser  & 10 & 10 & 25 & 480   \\ \hline
    \end{tabular}
    \caption{Statistics for each input model, showing the complexity in number of parts ($n_p$), number of connectors (\#C), number of connecting variations (\#CV), and number of design variations that define unique bag of parts $\mathcal{D}$.}
    \label{tab:modelstats}
\end{table}

\begin{figure}
    \centering
    \includegraphics[width = \linewidth]{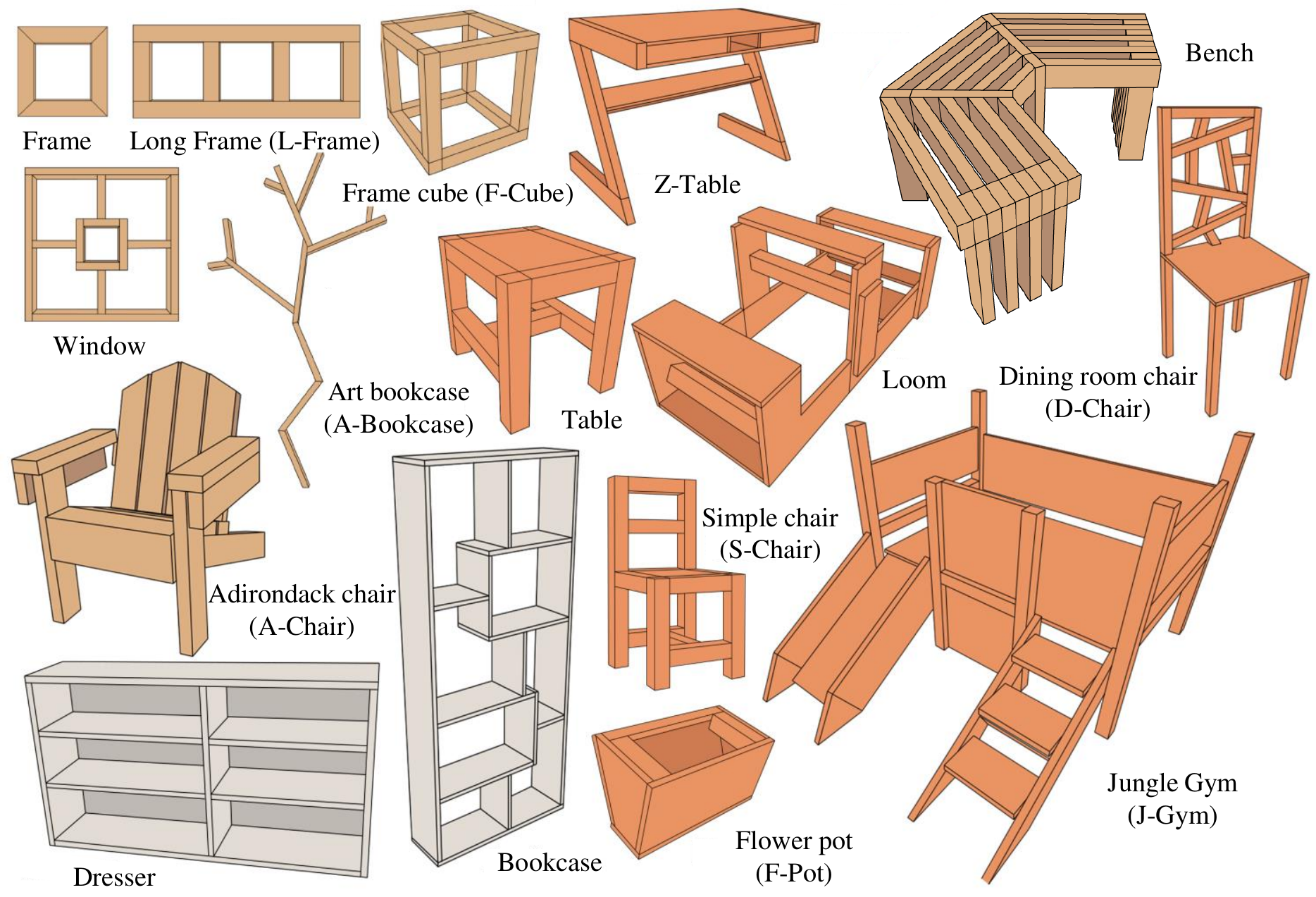}
    \caption{Models used for all experiments in \autoref{sec:results}. \change{Brown is used to indicate the models which are only made from 1D sequential cuts of lumber. Gray is for only from 2D partitioning of sheets. Orange is for both using 1D sequential cuts of lumber and 2D partitioning of sheets.  }}
    \label{fig:models}
\end{figure}

\section{Results and Discussion}
\label{sec:results}

In order to gauge the utility of our tool, we want to answer the following questions:
\begin{enumerate}


    \item How much does searching the design space with the fabrication space improve generated \fabplans?
    \item How does our tool compare with domain experts who are asked to consider different design variations?
    \item How does our tool's performance compare to a baseline na{\"i}ve approach?
\end{enumerate}

\subsection{Models}
We evaluate our method using the examples in \autoref{fig:models}. 
Statistics for each model are shown in Table~\ref{tab:modelstats}. 
\change{These models vary widely in visual complexity 
and materials used --- some are made from 1D sequential cuts on lumber,
where others require 2D partitioning of sheets.
Note the complexity of the search is not inherent to 
the visual complexity of the model,
rather, it is determined by the number of connecting variations
and the number of arrangements, 
which defines the size of 
the design space and the space of fabrication plans, respectively.
For example, the Adirondack chair is more visually complex
than the simple chair in \autoref{fig:models}, but because it has 
about 5000 times fewer design variations, it converges
much more quickly. 
Models of Art bookcase, Dining room chair, F-Pot, Z-Table, Bench, and Adirondack chair are taken from \cite{wu2019carpentry}. 
}

\subsection{Running environment}
\label{subsec:running-environment}

\change{

The parameters used in our ICEE algorithm are scaled based on the complexity of each model, measured in terms of the number of parts $n_p$ and the size of the design space $\nd$. We further introduce a single tuning parameter $\alpha \in [0.0,1.0]$, which allows us to trade-off between exploring more design variations (smaller values of $\alpha$) versus exploring more fabrication plans for given design variations (larger values). For all our experiments, we set $\alpha$ to the default value of 0.75. The ICEE parameters are set as follows: $K_d = 2^{\lceil { \log_{10} \nd }\rceil}$, $N_{pop} = 4 \cdot K_d$, $K_f = \beta \cdot n_p$, $K_{nd} = \lfloor (1.0-\alpha)\cdot K_d \rfloor$, and $P = 2 \cdot (\beta -2)$, 
$t_d = 10 $, $mt_d=200$, $mc_d = 0.95$, $mm_d = 0.80$, $T= 4$ hours, 
$t_p = 20 $, $mt_p=200$, $mc_p = 0.95$, $mm_p = 0.80$, $w = 0.7 $, $P_{rate}=0.3$ and $K_m = 10$,
where $\beta = \lfloor 44 \cdot \alpha^7 +2 \rfloor$. 

We report the running times of our algorithm in \autoref{tab:runningstats} for the models in \autoref{fig:models}.
The above times are measured on a MAC laptop computer with 16GB RAM and a 2.3 GHz 8-Core Intel Core i9 CPU.
More discussion of the running time is in the supplemental material.
}

\begin{table}[ht]
\footnotesize
    \centering
    \begin{tabular}{p{1.2cm}|p{0.3cm}|p{0.4cm}|p{0.5cm}|p{0.6cm}|p{0.5cm}|p{0.7cm}|p{0.5cm}|p{0.5cm}}
    \hline
    \textbf{Model} &\textbf{\#O} &\textbf{\#Iter} &\textbf{\#EDV} & \textbf{\#Arr} & \textbf{\#PDV} & \textbf{CEt(m)} & \textbf{Et(m)} & \textbf{Total(m)}  \\
\hline Frame      & 2 & 11 & 8   & 181    & 3  & 0.7  & 2.1   & 2.8   \\
\hline L-Frame    & 2 & 24 & 19  & 2818   & 3  & 2.1  & 6.1   & 8.2   \\
\hline A-Bookcase & 3 & 25 & 25  & 28700  & 3  & 20.5 & 228.6 & 249.0 \\
\hline S-Chair      & 2 & 15 & 136 & 35656  & 6  & 27.6 & 122.0 & 149.6 \\
\hline Table      & 2 & 18 & 50  & 9346   & 9  & 5.9  & 34.9  & 40.8  \\
\hline F-Cube     & 2 & 23 & 4   & 3499   & 3  & 1.4  & 4.0   & 5.5   \\
\hline Window     & 2 & 23 & 116 & 81026  & 4  & 32.8 & 98.9  & 131.7 \\
\hline Bench      & 2 & 25 & 16  & 37436  & 3  & 30.3 & 215.1 & 245.4 \\
\hline A-Chair    & 2 & 28 & 4   & 14440  & 3  & 3.1  & 9.6   & 12.7  \\
\hline F-Pot    & 3 & 14 & 3   & 185    & 2  & 1.7  & 13.0  & 14.7  \\
\hline Z-Table    & 3 & 70 & 41  & 336091 & 6  & 17.1 & 71.1  & 88.2  \\
\hline Loom       & 3 & 21 & 10  & 1812   & 5  & 3.1  & 74.6  & 77.7  \\
\hline J-Gym      & 3 & 46 & 18  & 286239 & 3  & 37.0 & 72.0  & 109.0 \\
\hline D-Chair    & 2 & 18 & 40  & 15054  & 7  & 27.7 & 228.8 & 256.5 \\
\hline Bookcase   & 3 & 15 & 32  & 34756  & 11 & 39.4 & 336.8 & 376.3 \\
\hline Dresser    & 3 & 20 & 44  & 22209  & 5  & 14.1 & 241.2 & 255.4
\\ \hline
    \end{tabular}
    \caption{
Some statistics and running times for our ICEE algorithm.
\change{For each model, we firs report the number of targeting objective (\#O) where 2 indicates material usage ($f_c$) and fabrication time ($f_t$), and 3 indicates  all of the three objective including cutting precision ($f_p$).}
We also report the number of iterations (\#Iter), explored design variations (\#EDV) and arrangements (\#Arr), and Pareto front design variations (\#PDV).
We report the running time of \bope contraction and expansion (CEt), and Pareto front extraction (Et), as well as the total time. All running time are in minutes.}
    \label{tab:runningstats}
\end{table}

\subsection{Benefits of Design Exploration}

\begin{figure*}
    \centering
    \includegraphics[width = 0.95\linewidth]{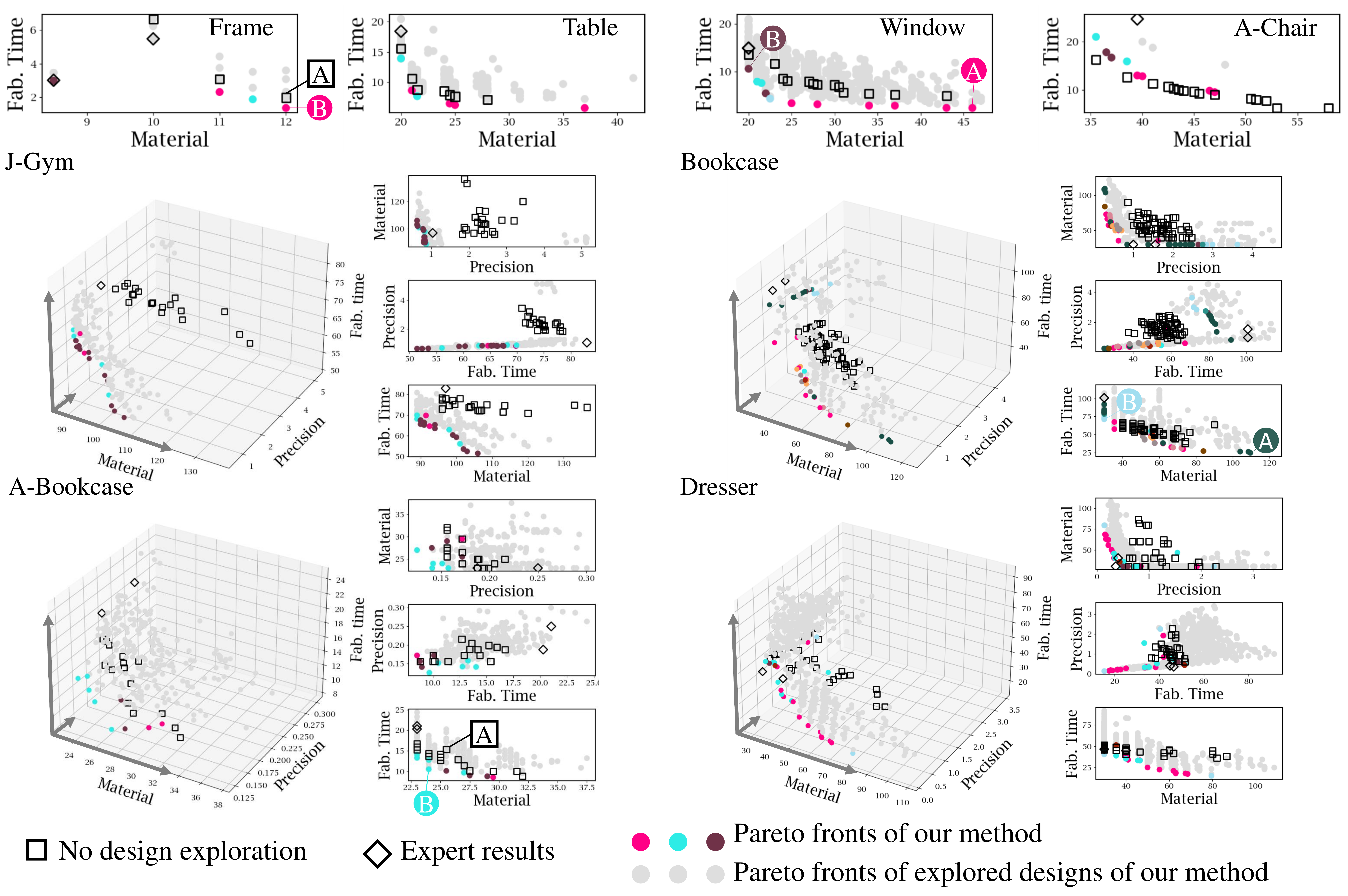}
    \caption{Pareto fronts computed from our pipeline with design optimization as colored dots. Each color corresponds to a different design. \change{The gray dots indicate the Pareto fronts of all explored design variations.} These are compared against Pareto fronts computed without design optimization (fabrication optimization only, using the original model as the input design) as squares, and expert fabrication plans  as diamonds. Often, fabrication plans from a design variant are more optimal than those generated from an input design. \change{For the unit of objective metrics, material usage ($f_c$) is in dollars, cutting precision ($f_p$) is in inches, fabrication time ($f_t$) is in minutes. Some (design, fabrication plan) pairs indicated with capital letters are visualized in~\autoref{fig:comparison-design-optimization-fab} and~\autoref{fig:comparison-design-optimization-fab-tradeoff}.}}
    \label{fig:comparison-design-optimization}
\end{figure*}

To demonstrate the benefit of simultaneous exploration of the design variation 
 and fabrication plan spaces, 
 we compare our tool against 
 optimizing the fabrication plan for a single design. 

\autoref{fig:comparison-design-optimization}
 shows the comparison between our pipeline and the Carpentry Compiler pipeline~\cite{wu2019carpentry}
 which only considers a single design.
 \change{The parameter setting of their pipeline and additional results can be found in Section 2 of the supplemental material.}
 We explore the trade-offs for fabrication time and material usage for the designs where all cuts can be executed with standard setups (these are considered to have \change{no precision error}) and include a third objective of precision error for the models where that is not possible.  
The Pareto fronts across designs generated by our tool cover a larger space and 
 many of our solutions dominate those from previous work. 

\change{
Exploring design variations enables better coverage of the Pareto front, which enables
finding better trade-offs. These trade-offs are lower-cost overall, cover more of the
extrema, and are more densely available. 
For example, a hobbyist may want to minimize material cost independent of time, as the manufacturing process is enjoyable, and they consider it to have no cost. Material cost is hard to save, but our exploration of design variations enable solutions that reduce material cost by 7\% in the Loom, 7\% in the Jungle Gym, 15\% in the Frame, and 25\% in the Bookcase. On the other hand, someone with free access to reclaimed wood may only care about the total manufacturing time. Our approach enables solutions that reduce fabrication time by 60\% --- two models saved between 50-60\%, three between 30-35\%, and four between 20-30\%, for example --- a huge time savings. If creating a very precise model is imperative, and a user would take great care to manufacture it exactly, then for four models, we find solutions that reduce error by 61-77\%. The detailed data are listed in Table S5 of the supplemental material.


Some examples don't lie at the extrema:
businesses often need to find a balance between the cost of materials,
time spent on a project, and overall project quality,
and the particular tradeoff will depend on their accounting needs.
Our method enables finding solutions with better tradeoffs.
Concretely, consider a carpenter charging \$40/h.
When scalarizing our multi-objective function into this single
objective of money,
we have several examples where the lowest cost on our Pareto front
is 5-8\% cheaper than the lowest cost on the baseline Pareto front,
such as the Z-Table, Flower pot, Jungle Gym, Dresser, Bookcase, and Art Bookcase.
The window and frame have cost savings of of 12\% and 20\%, respectively.
Though a cost reduction of several percent might appear insignificant,
in production at scale, it represents thousands of 
dollars potentially saved.
This scalarization function is just one way
for a user to judge the tradeoff between different
aspects of the multi-objective function.
In reality, the user probably has some notion of 
what tradeoff would be ideal for their purposes,
and will use the pareto front to
represent the full space of options and 
make an informed choice. This scalarized tradeoff is further examined in the Table S7 of the supplemental material.
}

\begin{figure}
    \centering
    \includegraphics[width = \linewidth]{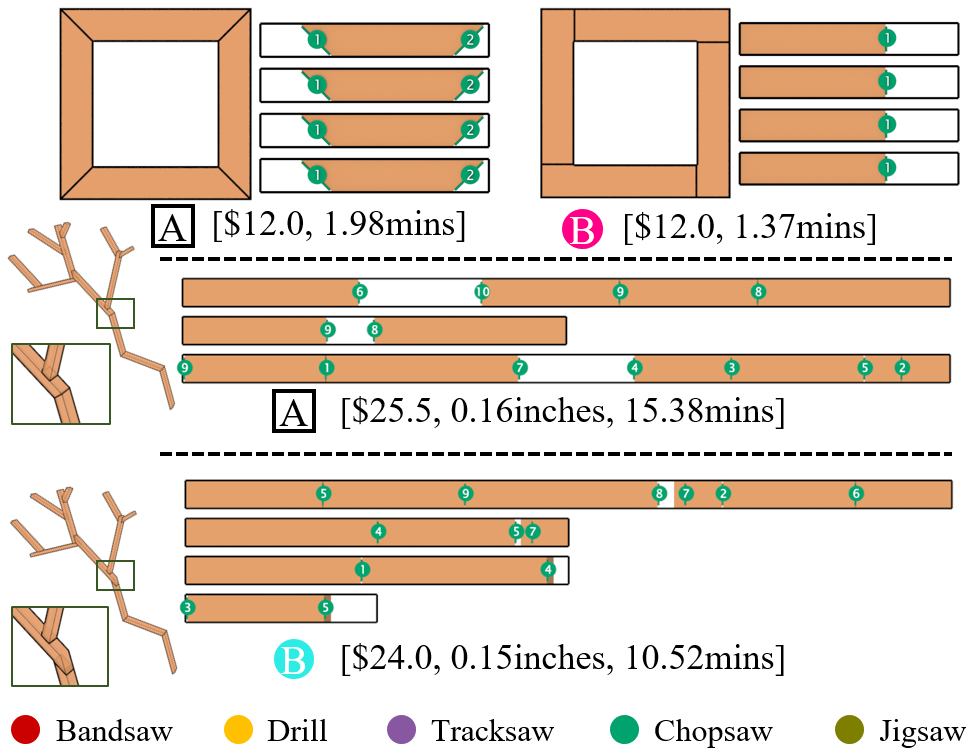}
    \caption{
    \change{
    Two examples where searching the design space revealed 
    fabrication plans that completely dominated the 
    fabrication plans generated for the input design.
    With the design variations, our pipeline could search for a design variation of the frame which turns all angled cutting to vertical. With Design B, we find a fabrication plan which takes less time than the least time-consuming plan A of the input design. Similarly, we show two fabrication plans of the A-Bookcase model where the design and fabrication plan B dominates the input design A. The fabrication costs are indicated in the figure with the order of material cost, precision error, and fabrication time. The cutting orders are labeled with colored dots and numbers, where colors indicate selected cutting tools, and stacked cuts are labeled with the same number.
    }
    }
    \label{fig:comparison-design-optimization-fab}
\end{figure}

\begin{figure}
    \centering
    \includegraphics[width = \linewidth]{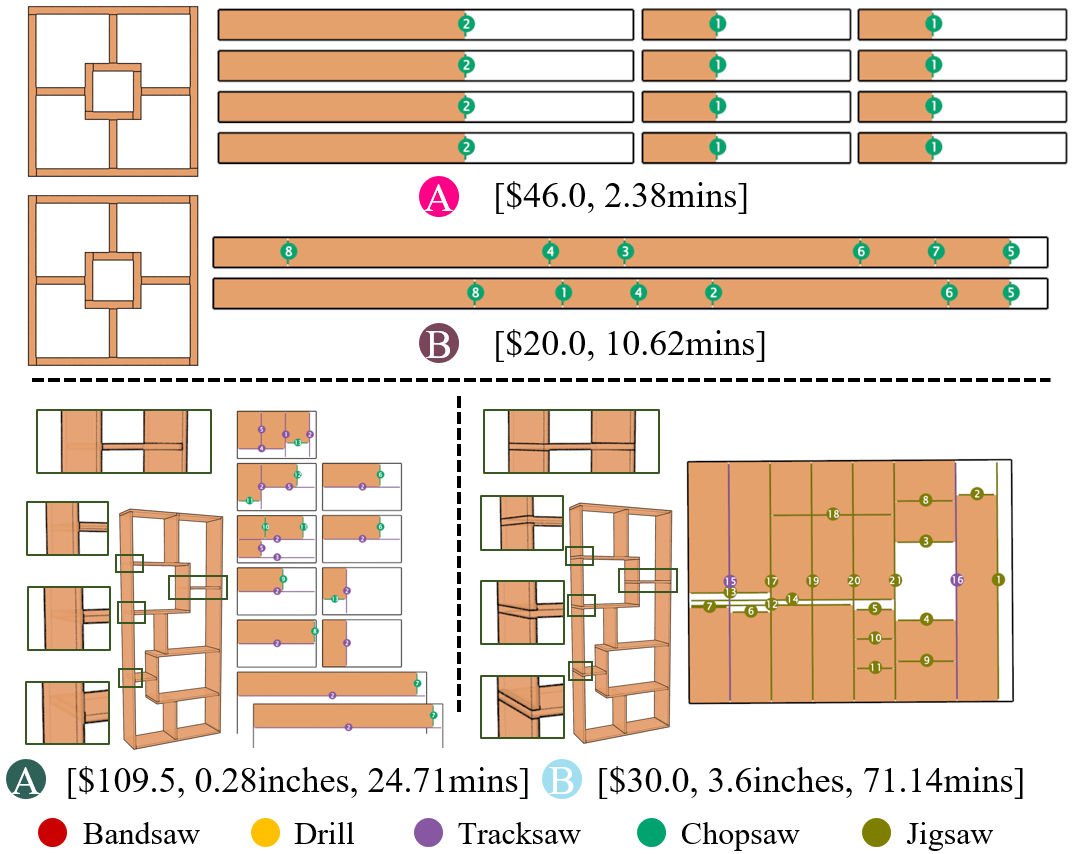}
    \caption{
    \change{Two examples where exploring different designs 
    lead to a wider diversity of plans,
    where each tradeoff on the Pareto front is 
    only possible because of the underlying design.
    The window provides a simpler example. 
    Design A is very uniform, with only three distinct
    parts. This design makes it easy to save on fabrication time
    because we can stack the cuts across different stocks.
    Design B features more varied cuts, unlike A,
    where each of the sides was the same length.
    This irregularity allows all the parts to be effectively
    rearranged onto just two pieces of stock. Regular pieces
    would not fit as nicely and result in wastage. Material 
    cost is very low, but because of the tight packing,
    much more time is needed to make each individual cut.
    The bookcase example showcases how some 
    unintuitive design decisions lead to cost savings.
    In this example, Design A's two long, identical
    side pieces mean more opportunities for stacking,
    of which the fabrication plan takes full advantage.
    This design enables a very low time cost, but 
    uses a lot of material. 
    Design B's left side is broken up by the shelves,
    and without a second long piece, it is possible 
    to pack all the pieces onto a single piece of lumber.
    Here, the material usage is economical, but 
    the carpenter must take time to
    cut pieces from a complex layout.}
    }
    \label{fig:comparison-design-optimization-fab-tradeoff}
\end{figure}

\change{
\autoref{fig:comparison-design-optimization-fab} highlights how exploring design variations generates fabrication plans that can dominate those
generated from no design variation exploration.
\autoref{fig:comparison-design-optimization-fab-tradeoff} then 
demonstrates how design variations enable diverse tradeoffs 
that save on different costs.
}

\subsection{Comparison with Experts}





For each model, we asked carpentry experts to generate design variations and \fabplans.
The resulting points are plotted as diamonds in \autoref{fig:comparison-design-optimization}.
Since experts produce each solution by hand,
 they produced Pareto fronts with many fewer solutions than our tool.
\change{For 14 of 16 models (except the Loom and Dresser models)},
solutions generated by our tool dominate the expert solutions.
This suggests that, generally, although expert plans seem sensible,
 our tool generates better output thanks to its 
 ability to generate more design variations and \fabplans, 
 including potentially unintuitive packing or cutting orders,
 and evaluate them much more quickly than a human.

\subsection{Performance Evaluation}
\label{subsec:perf-eval}

To test whether the \bope's sharing is important for our tool's performance,
 we compare against a nested-optimization pipeline built on the Carpentry Compiler~\cite{wu2019carpentry}.
The baseline approach invokes the Carpentry Compiler pipeline on
 each design variant that our tool explores,
 and then it takes the Pareto front across all design variations. 

\begin{table}
\footnotesize
    \centering
    \begin{tabular}{crr|rr}
     \multirow{2}{*}{Model} & \multirow{2}{*}{$\nd$} & \multirow{2}{*}{\#EDV} & \multicolumn{2}{c}{Time (min)} \\
    \cline{4-5}
     &  & & Ours & Baseline  \\
    \hline
{Frame}       & 13& 8  & 2.8 & 6.5 \\
{Jungle Gym}  & 54& 18 & 109.0 & 761.2\\
{Long frame}  & 65& 19 & 8.2 & 59.7\\
{Table}  & 1140& 59 & 40.8 & 612.8  \\
{Window}  & 10463 & 116 & 131.7 & 2050.0 
    \end{tabular}
    \caption{
    Results of the performance validation experiment.
    ``Ours'' indicates the ICEE algorithm of this paper. 
    ``Baseline'' indicates extracting the Pareto front \fabplans for each design variation explored by our method independently with the Carpentry Compiler pipeline~\cite{wu2019carpentry}. 
    The size of design space $\nd$ and the number of explored design variations (EDV) are also reported.
    \change{Our method and the baseline method produce 
two Pareto fronts which are indistinguishable.
This conclusion is not shown here; direct comparisons of hypervolume
can be non-intuitive due to the scale and how hypervolume is measured.
Please refer to the supplemental material (Figure S1), which contains plots
comparing the results of the two methods. 
Even with identical results, our time improvement is significant.}
}\label{tab:time-comparison}
\end{table}

\change{
We choose five models of varying complexity to evaluate performance and show results in \autoref{tab:time-comparison}.
 We tuned the parameters of the baseline method so we could achieve results that were
 as close as possible, if not qualitatively the same (when the baseline method ran to completion).
 Full results are available in the supplemental material (Table S6 and Figure S1).
This indicates that our co-optimization approach yields similar results to the nested approach over the same space.
When it comes to performance, 
 our approach is about one order of magnitude faster.
We attribute this speedup to the sharing captured by the \bope;
 we only had to evaluate common sub-designs and sub-fabrication-plans one time,
 whereas the baseline shared no work across its different invocations for each design variant.
}

\subsection{Fabricated Results}

\begin{figure}[h!]
    \centering
    \includegraphics[width = \linewidth]{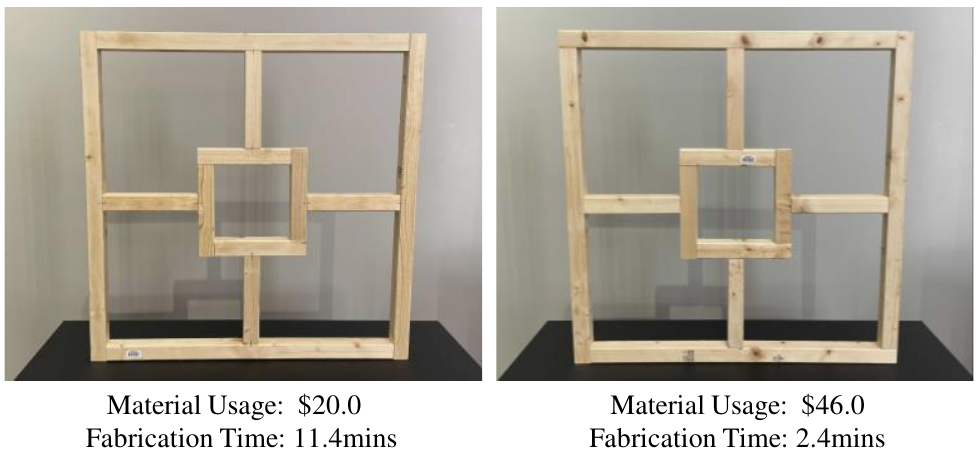}
    \caption{Fabrication results of two window variations. The different designs and fabrication plans trade off fabrication time and material usage.}
    \label{fig:fabrication-result}
\end{figure}

We validated our pipeline by physically constructing some of the models 
 according to the design variation-fabrication plan pairs generated by our tool.
\autoref{fig:fabrication-result} shows the results.

\section{Discussion}



\begin{figure}
    \centering
    \includegraphics[width = \linewidth]{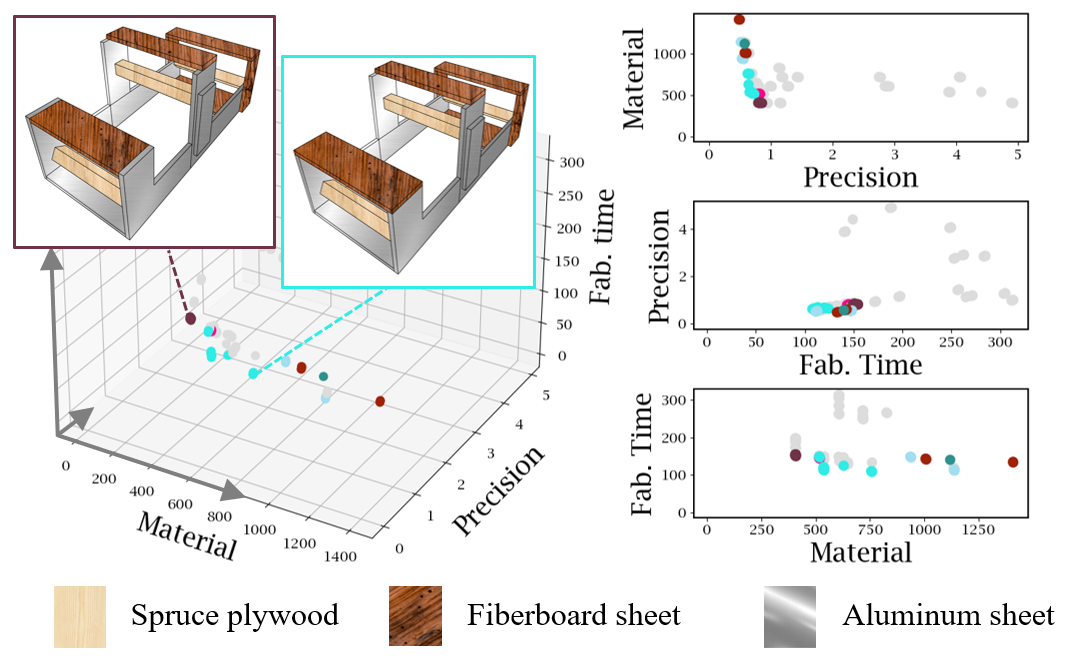}
    \caption{
    \change{
    A loom model with mixed material where two kinds of wood (spruce plywood and medium density fiberboard sheet) and one kind of metal (aluminum sheet) are assigned to each part.   
    }
    }
    \label{fig:mixed-material-Loom}
\end{figure}

\subsection{Multi-Materials and Cutting Tools}

\change{
Mechanical or aesthetic reasons might motivate designers to combine multiple materials, such as different types of wood, or wood and metal, in one model.
Adding new materials to our approach involves almost no implementation overhead:
we must select which cutting tools are appropriate, and accommodate
the material's costs into our metrics. Then, we simply need to indicate
which material a given part is made of, exactly the same way
we designate whether parts belong on 1D lumber or 2D stock.
As shown in \autoref{fig:mixed-material-Loom}, we have created a mixed-material model to showcase our ability to handle this added design complexity. The loom is made of two different types of wood as well as one kind of metal.
All parts are optimized in the same \egraph and 
treated identically to the base
implementation. We describe the cost metrics for different materials in the supplemental material (Section 1.3.1). 
}

\begin{figure}
    \centering
    \includegraphics[width = 0.95\linewidth]{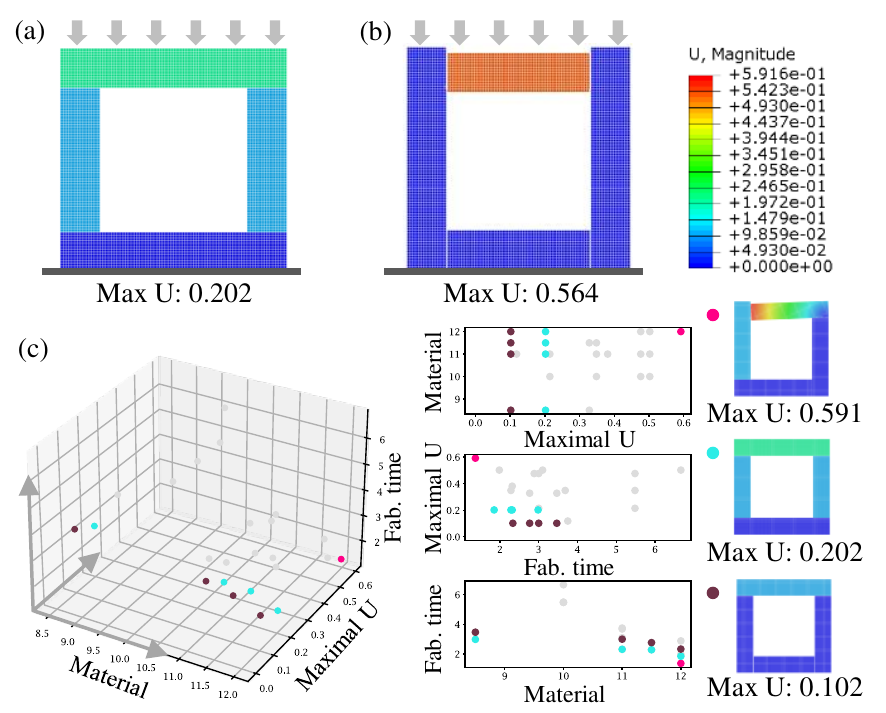}
    \caption{
    \change{
    Pareto fronts computed from by our pipeline for the Frame model with three objective functions, material usage $f_c$, fabrication time $f_t$ and stability performance. The physical stability of each design variation is simulated with Abaqus/CAE 2021, measured with the maximal displacement (Max U).  All displacements are in inches. In this figure, (a) is the displacement visualization in a direction, (b) is the displacement visualization of the same design but with a different direction, (c) plots the Pareto fronts computed from our pipeline where three design variations are selected.
    }
    }
    \label{fig:frame-cube-design-opt-comparison}
\end{figure}

\subsection{Objectives}


  
  



\change{
Our method also naturally extends to other objective functions.
We show one example in \autoref{fig:frame-cube-design-opt-comparison},
 where we consider stability as an additional objective
 which we calculate with physical simulation.
Notably, stability is invariant to the fabrication plan,
and depends solely on the design itself, so it only needs to be 
measured once, at the root node.
However, two designs can have different stability costs but share the same bag of parts.
\autoref{fig:frame-cube-design-opt-comparison} (a) and (b), 
exhibits one bag of parts which captures 
two different designs.

In this example, since the other metrics (time and material cost)
do not exhibit this dependency, we can simply assign to the root nodes
the stability cost of the best-performing design 
that corresponds to that bag of parts; thus
the cost for any given
bag of parts is the best cost of any design that is represented by
that bag of parts.
Note that fabrication plans depend solely on the bag of parts.
In general, if we want to use more than one metric like this one ---
a metric that depends on the design,
and is not completely determined by a term in the e-graph ---
we would need to compute the different trade-offs for the
variations during extraction, as was done 
with cutting order and precision, described in Section~\ref{subsec:extract}.


}





\subsection{Convergence} 

\change{While our results show the significance of the approach to reduce fabrication cost in practice, we cannot make any guarantees that plans we output are on the globally-optimal Pareto front. Indeed, we do not anticipate that any alternative approach would be able to have such strong guarantees given the inherent complexity of the  problem. This convergence limitation impacts our method in three different ways.

\paragraph{Parameter Tuning} Due to limitations in exploring the full combinatorial space, parameters of our search algorithm may influence convergence. Because the key aspect of ICEE is simultaneously searching  ``broad'' (design variations) and ``deep'' (fabrication plans for various designs), we expose the $\alpha$ parameter that trades-off between depth and breadth during search. Exposing this single parameter, enables us to improve performance in special circumstances. For example, when not much can be gained from design variations, a larger $\alpha$ will enable searching deeply on a particular design finding better solutions. All the results shown in this paper use the default value for $\alpha$ that we have found effective in practice.

\paragraph{Comparison with~\citet{wu2019carpentry}} The fundamental difference between our work and~\citep{wu2019carpentry} is that incorporating more design variations increases the design space, enabling us to find better performing results. Since the search space of this prior work is a subset of the search space we explore, our results should be strictly better. However, since neither method can ensure the results lie on the true Pareto front due to limitations in convergence, tuning parameters of both approaches may influence this result. An example of this limitation in shown in A-Chair example in Fig7. We show in the supplemental material (Section 2.4) how tuning $\alpha$ to explore more deeply improves this result and also report experiments for tuning the 4 parameters from~\citep{wu2019carpentry}. 

\paragraph{Increasing the Design Space} A final implication of the intractable search is that it is possible to achieve worse results by increasing the design space in special circumstances. We discuss in the supplemental material (Section 2.4) an example where we make the design space 145 times larger by including variations that do not benefit the search. 
}

\subsection{Limitations and Future Work}

\label{subsec:lim}
 Our current approach encodes only discrete design
 variants in the \bope.
An interesting direction for future work would be to
  support continuous variations in the designs space
  which can provide a larger space of fabrication plans
  to explore.
However, supporting continuous design variants in an \egraph
 would require designing a new metric for comparing two
 design variants for equivalence.
This is challenging because e-graphs
 heavily exploit transitivity,
 so any error in the metric
 could lead to arbitrarily different designs being
 considered ``equivalent''.

Several steps of our algorithm can
  also be parallelized for performance
  (e.g. generating design variants)---we leave this
  as an extension for the future.

We are also keen to explore broader applications of the
  ICEE strategy for integrating feedback-directed search
  in other \egraph-based optimization techniques.
Past work applying \egraphs for design optimization
  in CAD~\cite{szalinski} and for
  improving accuracy in floating-point code~\cite{herbie}
  have relied on ad hoc techniques for controlling the
  growth of the \egraph, e.g., by carefully tuning
  rules used during rewrite-based optimization.
We hope to explore whether ICEE can help
  with optimization in such domains by
  focusing the search on more-profitable candidates
  and easing implementation effort by reducing the
  need for experts to carefully tune rewrite rules.

\change{The most time-consuming part of our ICEE algorithm lies in the Pareto front extraction phase. A pruning strategy with learning-based methods for predicting the objective metrics of an arrangement might be an interesting and valuable area of research.}

Another direction we are eager to explore is
  accounting for other factors in the Pareto front.
Currently, our technique finds a
 design variant and fabrication plan
 that optimizes fabrication time,
 material cost, and
 precision.
Other interesting factors that can guide the
 search include ease of assembly
 and strength of the part.

\section{Conclusion}

We have presented a new approach to co-optimizing model design variations and their fabrication plans.
Our approach relies on the insight that 
 fabrication plans across design variants will share similar structure.
We capture this sharing with the \bope data structure 
 that considers fabrication plans equivalent if they produce the same bag of parts.
The \bope also lets us guide the search toward profitable design 
 variants/fabrication plans with a technique we call ICEE 
 (Iterative Contraction and Expansion of \egraphs)
 that may be useful for uses of \egraphs in other applications.
Results generated by our tool compare favorably against 
 both expert-generated designs and a baseline built using prior work,
 indicating that the sharing captured by the \bope is essential 
 to efficiently exploring the large, combined space of design variants and fabrication plans.

\bibliographystyle{ACM-Reference-Format}
\bibliography{sample-bibliography}


\end{document}